\documentclass[10pt,letterpaper]{article}
\usepackage{opex3} 

\begin{document}

\title{Multiplexed broadband beam steering system utilizing high speed MEMS mirrors}

\author{Caleb Knoernschild$^{1*}$, Changsoon Kim$^{1}$, Felix P. Lu$^{2}$, and Jungsang Kim$^{1}$}

\address{$^{1}$Duke University, Electrical and Computer Engineering Department, Durham, NC 27708\\$^{2}$Applied Quantum Technologies, Durham, NC 27707}

{\vskip-.3cm \parskip0pc\hskip2.25pc \footnotesize%
   \parbox{.8\textwidth}{\begin{center}\it $^*$Corresponding author: \textcolor{blue}{\underline{caleb.k@duke.edu}} \rm \end{center} } \normalsize  \vskip-.2cm}



\begin{abstract*}
We present a beam steering system based on micro-electromechanical systems technology that features high speed steering of multiple laser beams over a broad wavelength range. By utilizing high speed micromirrors with a broadband metallic coating, our system has the flexibility to simultaneously incorporate a wide range of wavelengths and multiple beams. We demonstrate reconfiguration of two independent beams at different wavelengths ($780$ and $635$ nm) across a common 5$\times$5 array with $4$ $\mu$s settling time. Full simulation of the optical system provides insights on the scalability of the system. Such a system can provide a versatile tool for applications where fast laser multiplexing is necessary.\\ \\
\end{abstract*}


\bibliographystyle{osajnl} 


\section{Introduction}
\label{intro}

Efficient utilization of laser resources by controllably directing or steering light from a single source across a relatively large area is a topic that affects a wide variety of research interests. Studies in imaging \cite{ShinOE2007,Jung2005,RollinsOE1998}, optical communication networks \cite{KimIPTL2003,AksyukLTJo2003}, optical data storage devices \cite{KimSAAA2004}, and projection display technologies \cite{VanPotI1998,ConantSAAA2000} all make use of beam steering devices to improve system effectiveness. Many of these systems are implemented using mechanical structures such as galvanometer mirrors or microelectromechanical systems (MEMS). The steering speeds of these systems have been limited to tens of kilohertz or less due to the mechanical resonant frequencies of the steering mirror elements.

There are still other applications that require steering speeds approaching the MHz range. Some of these examples include atomic based quantum computing where the time to reconfigure the position of the laser (settling time) must be less than the decoherence time of a quantum state stored in a single ion or neutral atom (qubit) \cite{LeibfriedN2003,Schmidt-KalerN2003,SaffmanPRA2005}. Because of the high speed requirements for these experiments, acousto-optical or electro-optical deflectors are commonly used to provide the steering function. While these strategies have an advantage in speed, their limitations present significant obstacles. Both acousto-optical deflectors and electro-optical deflectors have to be wavelength tuned and require complex engineering to incorporate multiple wavelengths \cite{KimAO2008}. In addition, acousto-optical deflectors need $\sim1$ W RF drive power and induce small frequency shifts in the laser that must be accounted for, while electro-optical deflectors need large operation voltages and have limited angular range. Both technologies are generally restricted to single beams and scaling to a large number of independent beams is not straightforward.

We previously reported our implementation of a MEMS based 2 dimensional (2D) single beam steering system \cite{KnoernschildOL2008}. In this paper, we demonstrate the scalability of the system by incorporating two beam paths at different wavelengths ($780$ nm and $635$ nm) with substantial improvements in steering speed and optical throughput compared to our previous results. This system utilizes highly optimized MEMS mirrors and features scalability to multiple beams while achieving settling times as low as $4$ $\mu$s. In order to investigate the scalability of the system to larger numbers of beams, we performed optical modeling of the full system. Our approach can easily accommodate multiple independent beams over a wide range of wavelengths and controllably direct them to any random position within a $5\times5$ array. Furthermore, each beam path can be arranged to deliver multiple wavelengths of light simultaneously. In this paper we discuss the optical system design, MEMS mirror design, simulations used to investigate the scalability of the system, and the results of a two laser beam steering system.


\section{System Description}
\label{Sys_Descrip}

Certain atom-based quantum information processing (QIP) applications require reconfigurable beam paths for multiple wavelengths of lasers to perform logic gate operations. For example, a two qubit gate operation utilizing dipole-dipole interactions of Rydberg states in trapped $^{87}$Rb atoms requires $780$ and $480$ nm lasers to excite (de-excite) atoms between the Rydberg and ground states \cite{JohnsonPRL2008}. We engineer our system to meet the requirements of such an experiment. The baseline operation of the design will direct two independent laser beams at different wavelengths to 25 different lattice sites in a 5$\times$5 array. Each site must be individually addressed by the beam with minimal residual intensity at neighboring sites. The separation between adjacent lattice locations, dictated by the boundary conditions in the atomic physics experiment, is defined to be $a=10$ $\mu$m with a beam waist at the lattice of $w_o=a/2=5$ $\mu$m, and the system must shift the laser a full beam diameter ($2w_o=a$) to the neighboring lattice location. Therefore, the extent of the steering range requires $\pm4w_o$ in both dimensions. Due to qubit decoherence times it is desirable that the system steers the beam among lattice sites in a few microseconds. 

\begin{figure}[htb]
\centering
\includegraphics[width=12cm]{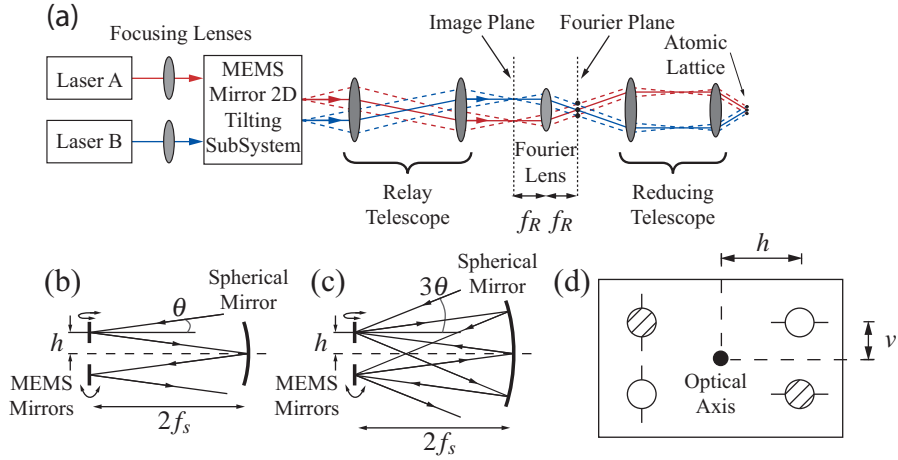}
\caption{\label{fig:schematic} (a) Schematic of a two beam steering system. (b) 2$f$-2$f$ folding imaging optics to combine decoupled tilt motion for a single bounce system ($n=1$). The dotted line indicates the optical axis. (c) Double bounce system ($n=2$). (d) Top view a MEMS device for a two beam layout. One beam utilizes the solid white mirrors while the second beam uses the patterned mirror set. Horizontal and vertical separations from the optical axis are labeled as $h$ and $v$ respectively.}
\end{figure}

Fig.~\ref{fig:schematic}(a) shows the schematic for a two beam steering system. Our system design uses MEMS mirrors to provide 2D tilting of the beams while a lens (Fourier lens) converts the tilts into displacements on the plane of the target array. Relay optics create sufficient room for placement of Fourier lens without clipping the beam path. Additional telescope projection optics after the Fourier lens reduce the beam waist to the size required at the lattice. Because the steering is accomplished by a reflective element, the system can operate in a wide range of wavelengths, which is only restricted by the coating on the MEMS mirrors and other optical components. A reflective steering design also enables wavelength multiplexing by aligning multiple wavelengths along the same beam path. Matching the Rayleigh lengths of each wavelength in this case ensures consistent imaging of the beam waist. Utilizing MEMS technology allows our system to have the scalability to address larger arrays and multiplex multiple beams onto the same array.

The core of the system design comes from the 2D tilting subsystem. While a single MEMS mirror that provides 2D beam steering has been demonstrated \cite{SuIPTL2001}, it is difficult to reach the target speeds with such mirrors. Using a pair of small one dimensional (1D) tilting MEMS mirrors with limited angular range \cite{KnoernschildOL2008} we can achieve significantly faster beam steering performance than the 2D mirrors. In order to accommodate two axis motion for a single beam path, two 1D mirrors are oriented with orthogonal rotational axes horizontally separated by $2h$ on the same substrate. A spherical mirror with focal length $f_{s}$ in a folded $2f$-$2f$ imaging configuration is used to direct and focus the reflection from the first mirror onto the second thus combining the two orthogonal tilts as demonstrated in Fig.~\ref{fig:schematic}(b). In terms of the Gaussian beam, the imaging improves to first order as $\left(z_{R}/f_s\right)^2\rightarrow0$ where $z_R$ is the beam's Rayleigh length.

To compensate for the device's limited angular range, a system level angular multiplication scheme is used. The incoming laser's incident angle {$(2n-1)\theta$} induces $n$ reflections off each MEMS mirror (Fig.~\ref{fig:schematic}(c)), where $\theta \approx h/2f_s$ is the incident angle that sends the reflection from the first mirror to the center of the spherical mirror. Multiple reflections ($n>1$) increase the subsystem's angular range to $2n\phi$ for a given MEMS mirror mechanical tilt angle $\phi$. Increasing $n$ to produce more dramatic angular multiplication requires a larger incident angle, and the beam paths experience larger aberrations in the $2f$-$2f$ imaging process. Furthermore, the optical throughput is reduced when mirror reflectivity is below unity. We employ a double reflection system ($n=2$) to provide twice the angular range of a single reflection system while maintaining adequate control over optical system aberrations and throughput.

Because the telescope projection optics can be used to reduce the beam waist down to the necessary value, the radius of the MEMS mirror can be used as a design parameter to increase steering speed. Based on the radius of the mirror (discussed in the next section), a beam waist of $\approx40$ $\mu$m at the MEMS mirror is chosen, and a spherical mirror with $f_s=50$ mm (roughly 10 times the Rayleigh length) is used for a compact system while maintaining adequate Gaussian beam imaging. To make the system easier to characterize, we use a $20$ mm focal length Fourier lens and a demagnifying relay telescope with $100$ mm and $50$ mm focal length lenses. This produces a beam waist at the Fourier plane of $250$ $\mu$m. In order for the edge mirror to completely capture the full range of the beam paths without clipping in the 2D tilting subsystem, we separate the mirror on the chip by $2h=9$ mm.

Scaling the system from a single to multiple beam paths can be achieved by a simple modification to the 2D tilting subsystem. Multiple pairs of MEMS mirrors located on opposite sides of the spherical mirror's optical axis can provide individual reconfiguration of each beam path. Fig.~\ref{fig:schematic}(d) shows the mirror arrangement for a two beam system on a single planar device. Two pairs of mirrors (one pair for each beam path) are symmetrically located about the optical axis with a vertical and horizontal offset, $v$ and $h$, respectively. The system is aligned such that all beam paths leaving the unactuated 2D tilting subsystem travel parallel to the optical axis with vertical offset $v$ and zero horizontal offset. After the relay telescope, these parallel beams are focused on axis at the Fourier plane by the Fourier lens. Tilts introduced by the MEMS mirrors break the parallel beam propagation and cause the paths to be shifted to a different positions at the Fourier plane. In order to maintain consistent propagation of the Gaussian beam across multiple colors, the Rayleigh length of each wavelength must be matched.

As the number of independent beams within the system increases, the physical arrangement and number of mirror pairs will force $v$ to increase for the outer most mirrors. These beam paths experience larger aberrations in the 2D tilting subsystem as well as the remaining optics as the offsets increase. The resulting aberrations lead to imperfections in the 2D tilting and relay subsystems which will be the subject of discussion in section \ref{sims}.


\section{MEMS Mirrors}
\label{mems}

\begin{figure}[htb]
\centering
	\includegraphics[width=10.6cm]{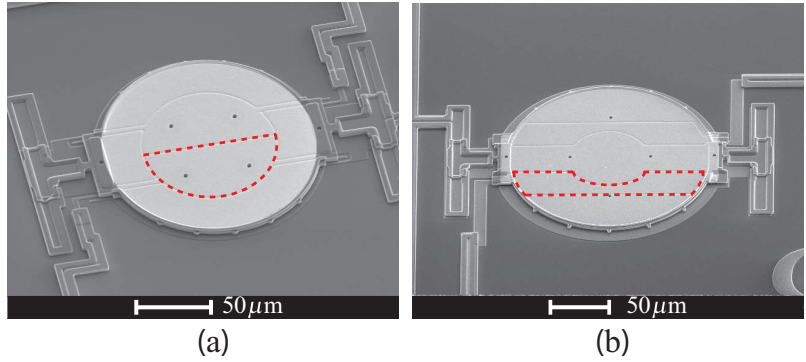}
	\caption{Scanning electronic micrographs with electrode print-through shape indicated by dotted red line. (a) shows an image of $75$ $\mu$m radius mirror. This mirror has a reflective gold coating and ``D''-shaped mirror electrodes. (b) shows an image of a $90$ $\mu$m mirror with ``U''-shaped electrodes.}
	\label{fig:combsem}
\end{figure}

The mirror design is strongly coupled to the optical system described above, driven by the settling time in the application under consideration. Since the target settling time of a few microseconds is significantly smaller than other reported steering systems, the mirror geometry has to be optimized for speed. The mirror design consists of a circular mirror plate rotating about 2 torsional springs (Fig.~\ref{fig:combsem})~\cite{ChangsoonSTiQEIJo2007}. Tilt is induced by electrostatic actuation from a voltage applied between the grounded mirror plate and underlying electrodes. The mirror can be rotated in a positive or negative direction, so the mirror's mechanical tilt only needs to address 2 of the 5 lattice sites in each direction away from the center position. The optical system design places requirements on the maximum mechanical tilt angle of the mirror ($\phi_{max}$) in relation to its radius ($r$). In our system, simple ray tracing and Gaussian beam considerations lead to the relationship between mirror radius and required mirror tilt angle, $r\propto1/\phi_{max}$ \cite{KnoernschildOL2008}. 

The dynamic characteristics of the mirror's torsional motion is described by the damped harmonic oscillator equation

\begin{equation}
\label{equ:dynamic}
 \ddot{\phi}\left(\tau\right)+2\zeta\dot{\phi}\left(\tau\right)+\phi\left(\tau\right)=\frac{1}{2I\omega^2_R}\frac{\partial C\left(\phi\right)}{\partial\phi}V^2\left(\tau\right).
\end{equation}
Here, $\phi$ is the mechanical tilt angle of the mirror, $\zeta$ is the damping ratio, $\tau=\omega_Rt$ is a dimensionless time variable, $V$ is the applied voltage between the mirror plate and actuation electrode, and $\omega_R=\sqrt{2\kappa /I}$ is the resonant frequency of the mirror where the springs have torsional stiffness $\kappa$ and the circular mirror plate has a moment of inertia $I \propto r^4$. In order to achieve the targeted transition speed, the settling time of the mirror's step response is minimized by increasing $\omega_R$ while maintaining near critical damping. Maximizing beam steering speed while meeting system requirements is a complex optimization process \cite{ChangsoonSTiQEIJo2007} due to physical limitations on available torsional stiffness $\kappa$ and control voltage \cite{Wallash2003} as well as the strong dependence of the damping ratio $\zeta$ on the mirror radius and the air gap under the mirror plate. We have found that mirrors with a radius close to $75$ $\mu$m and a gap of $1.25$ $\mu$m give the best compromise of resonant frequency and proper damping. For mirror designs with a gap of $2$ $\mu$m, a radius of about $100$ $\mu$m is ideal.

The mirrors are fabricated using the PolyMUMPS foundry process through MEMSCAP, Inc \cite{Memscap}. This process consists of one electrical routing layer and two structural layers of polysilicon. We utilized the electrical routing layer to create the actuation electrodes while the two structural layers were stacked to form the springs and mirror plates. Because of the conformal deposition process used for the polysilicon layers, the electrode pattern is printed through onto the mirror plate resulting in non-idealities on the reflecting surface. Etch holes on the mirror plate are also required to effectively remove all the sacrificial oxide which provides layer spacing for the device. Fig.~\ref{fig:combsem} shows scanning electron micrographs of typical MEMS mirrors in the system. To provide high reflectance for the mirrors, we deposit a thin layer of metal depending on the target operation wavelength range. Cr/Au is used for infrared wavelengths and aluminum is used for ultraviolet and visible wavelengths. Because of the mirror thickness ($3.5$ $\mu$m), it becomes difficult to introduce multi-layer dielectric coatings without significant stress engineering to maintain a flat mirror.

The thin mirror plate requires proper stress control of the metal reflective coating to maintain a flat surface. For a gold reflector, deposition with an initial seed layer such as chromium (Cr) or titanium (Ti) is common. While Ti can be placed on the device with much lower stress, the MEMS device processing requires a ``release'' step using a hydrofluoric acid (HF) etch after metal deposition, which removes the Ti. Cr etches much more slowly in HF but has significant intrinsic tensile stress that increases with thickness \cite{PicardBAB2002}. We used a very thin evaporated Cr layer ($\sim25$ \AA) with slow deposition rates ($0.5$ \AA $/$s) to minimize stress. With this process, we achieved mirrors with radius of curvatures of $30$ cm or better. 

The etch holes and electrode print-through patterns on the mirror plate induce scattering of the reflected light. Optimized design of electrode shapes and etch hole locations can minimize these effects. Because the majority of the optical intensity of a Gaussian beam will be incident on the center of the mirror, the etch holes are pulled away from the middle of the mirror plate. Moving the electrodes and therefore the print-through patterns further away from the center of the mirror plate reduces beam diffraction from these structures. Two different electrode geometries were fabricate and tested for their optical performance. In the first design, two ``D''-shaped electrodes create a print-through that travels along the rotational axis and intersects the center of the mirror plate (Fig.\ref{fig:combsem}(a)). The second design moves the print-throughs out of the center region with two ``U''-shaped electrodes (Fig.\ref{fig:combsem}(b)).

	
\section{Simulations}
\label{sims}

We used commercial ray tracing software (Zemax) to simulate the optical system, which provides quantitative analysis on the aberrations of the imaging subsystems, optical losses associated with clipping at the MEMS mirrors, and limitations on the scalability due to aberrations. The simulation of the system is broken down into three parts: the 2D tilting subsystem, relay telescope and Fourier lens, and the combined system. The model allows examination of spot diagrams, aberration diagrams, and Gaussian beam intensity plots. The MEMS mirrors are modeled as ideal reflectors without etch holes or print through patterns. Data from simulations have provided essential feedback for MEMS mirror placement on chip, lens selection, and design of custom compensation optics.

\begin{figure}[htb]
\centering
\includegraphics[width=12cm]{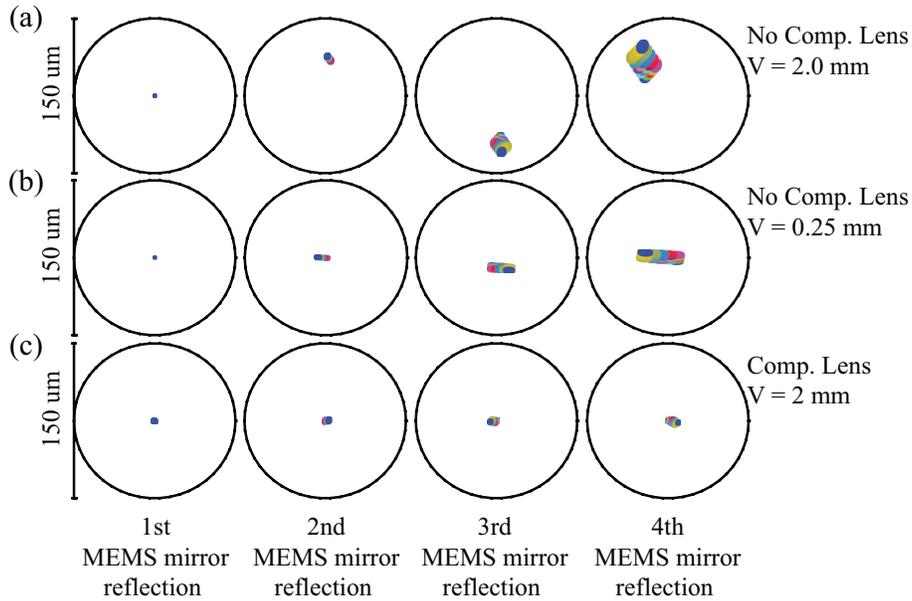}
\caption{\label{fig:spotDiagram} Plot shows spot diagrams for a single wavelength at each MEMS mirror reflection (columns) in the 2D tilting subsystem. Each mirror tilt configuration is plotted on the same set of axis and displayed in different colors. The circle represents a $75$ $\mu$m radius MEMS mirror. (a) System with $v=2$ mm with maximum decentering of $57$ $\mu$m. (b) System with $v=0.25$ mm with maximum decentering of $27$ $\mu$m. (c) System with $v=2$ mm and compensation lens with maximum decentering of $10$ $\mu$m.}
\end{figure}

The modeling begins with the 2D tilting subsystem. Since this subsystem is entirely made of reflective optics, the chromatic aberrations arising in a multi-wavelength system do not exist. The vertical and horizontal separation of the MEMS mirrors gives rise to Seidel aberrations that increase as the mirrors are tilted and are compounded by the angular multiplication scheme. Spot diagrams taken with 780 nm light at each MEMS mirror reflection for $h=4.5$ mm and $v=2$ mm are shown in Fig.~\ref{fig:spotDiagram}(a) where the spot diagrams from all 25 different tilting configurations are shown on the same set of axes. As the beam propagates through the folded imaging configuration, the reflections stray further from the center of the MEMS mirror (indicated by the $75$ $\mu$m radius circle) and the separation among the spot diagrams of the different mirror tilt configurations increases. This aberration decenters the beam which results in clipping on the MEMS mirror aperture as the system addresses different lattice sites. We can minimize the decentering by reducing the vertical offset $v$ of the MEMS mirrors. The horizontal offset $h$ is necessary to bring the beams in and out of the system, and cannot be reduced once the focal length $f_s$ of the imaging system and the angle multiplication factor $n$ are chosen. Reduction in $h$ must accompany reduction in $f_s$ and the impact on aberration does not significantly improve. Fig.~\ref{fig:spotDiagram}(b) shows the spot diagrams for a system where $v$ is reduced to $0.25$ mm. When the incident beam is aligned at the center of the first MEMS mirror, the maximum decentering at the last reflection on the MEMS mirror goes from $57$ $\mu$m for $v=2$ mm to $27$ $\mu$m for $v=0.25$ mm. Further improvement can be obtained by designing custom optical elements to correct the aberrations. Fig.~\ref{fig:spotDiagram}(c) shows the spot diagrams from a system with $v=2$ mm mirror separation that includes a custom aspherical lens (compensation lens) located just before the MEMS mirrors to compensate for the off-axis aberrations. This lens has one convex aspherical surface (radius of curvature of $38.53$ mm and a conic constant of $-19.44$) and one concave spherical surface (radius of curvature of $40.0$ mm) resulting in a maximum decentering of $10$ $\mu$m. While the compensation lens can improve the imaging quality, we chose not to implement it in our system for simplicity.

Since the beams leaving the 2D tilting subsystem may have a vertical offset $v$ and contain multiple wavelengths, proper design of the relay telescope and Fourier lens is essential to reduce off-axis and chromatic aberrations. We used off-the-shelf achromatic doublets to manage these aberrations for a two beam system at $780$ nm and $635$ nm. Combining the model for the relay and Fourier optics with 2D tilting subsystem model, we were able to perform a complete system simulation using Zemax's physical optics propagation feature. A $780$ and $635$ nm Gaussian beam source with a $37$ $\mu$m beam waist located at the first MEMS mirror was propagated through the entire system for each of the 25 different mirror configurations. The beam profile was examined at the Fourier plane to verify proper addressing of the $5\times5$ array, and peak intensity as well as total optical power data among all the configurations were examined. These simulations are repeated for a range of vertical MEMS mirror offsets and mirror radii.

The beam uniformity across the lattice sites is characterized by the peak intensity variation. We can isolate the effects of system's optical aberrations on the intensity variation by enlarging the MEMS mirrors to eliminate clipping in the 2D tilting subsystem. Because of the horizontal offset ($h=4.5$ mm) of the MEMS mirrors, the folded imaging subsystem slightly alters the Gaussian beam properties across the lattice sites. This results in $1\%$ peak intensity variation horizontally across the array. While the vertical offset $v$ of the MEMS mirror in the 2D tilting subsystem also generates similar intensity variations in the array, dominant contribution for the intensity variations across vertical locations arises from the aberrations in the relay and Fourier optics. Since the system is aligned such that the beam path is offset by $v$ from optical axis running through the relay telescope and Fourier lens, the vertical intensity variations increase as $v$ increases. The mirror configurations that force the beam paths further from the optical axis create smaller beam widths (and therefore larger peak intensities) at the Fourier plane compared to those paths closer to the optical axis. With a vertical separation of $v=2$ mm, the entire lattice features peak intensity variations of $7\%$ while a separation of $v=0.25$ mm results in only $1.7\%$ peak intensity variations.

As the size of the MEMS mirrors are decreased, the effects of aberrations in the 2D tilting subsystem cause clipping of the beam on the mirrors. The amount of optical power lost varies for different mirror tilt configurations. While this clipping induces little beam distortion at the Fourier plane, it causes larger variation of the peak intensity than that due to off-axis aberrations. By shifting the reflection point of the beam path away from the center of the first MEMS mirror, the spot diagrams from the $2^{nd}$, $3^{rd}$, and $4^{th}$ mirror reflections can be shifted closer to the center of the respective mirror. This minimizes beam clipping and reduces the peak intensity variation across lattice sites. For $75$ $\mu$m radius MEMS mirrors, a vertical offset of $v=2$ mm produces peak intensity variation of $12\%$ at the Fourier plane while $v=0.25$ mm reduces that number to $3.5\%$. 

Simulations indicate that our current system design can easily support 9 pairs of $75$ $\mu$m radius mirrors (9 beams) aligned in two columns on the substrate with $12\%$ or better peak intensity variation for each beam without any custom compensation optical elements. Introducing additional beams requires more pairs of mirrors which increases $v$ and therefore the peak intensity variations across the output array for the outermost mirrors. Changing the dimension of the mirrors, optimizing their placement, or using custom optical elements to compensate for aberrations can increase the number of beams the system can accomodate with minimal peak intensity variations.

	
\section{System Performance}
\label{results}

\begin{figure}[htb]
\centering
\includegraphics[width=12cm]{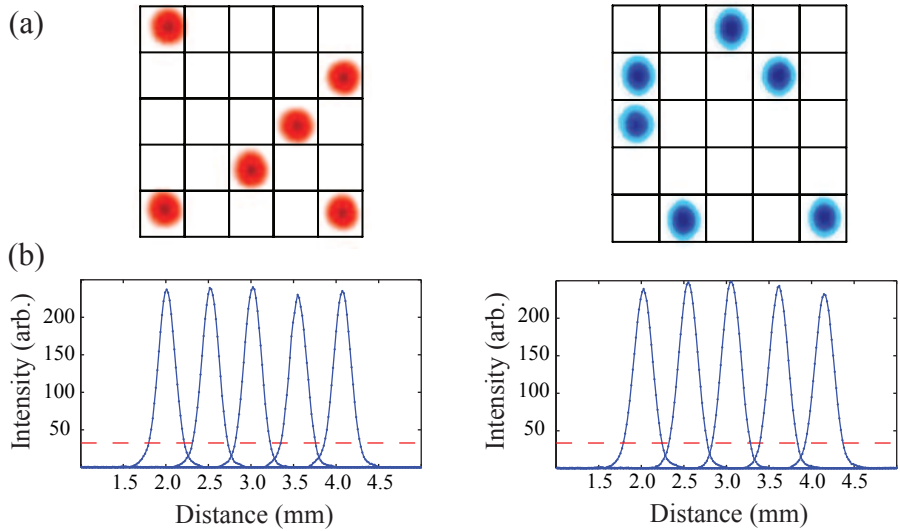}
\caption{\label{fig:profile} Gaussian beam profile data for $635$ nm (left) and $780$ nm (right) in a system with vertical mirror separation of $v=0.25$. (a) Random beam intensity data overlaid on the same set of axes showing addressability of the $5\times5$ array. (b) Intensity profile data indicating a complete beam waist shift to neighboring location. The dashed red line indicates the $1/e^2$ level.}
\end{figure}

For the functional demonstration of a two beam system, we used separate wavelengths of $780$ and $635$ nm with $80$ nm gold reflective coating to improve system throughput. Figure \ref{fig:profile} shows the Gaussian beam data collected for a system with a vertical mirror offset of $v=0.25$ mm. The top plots show beam intensity data taken for several different locations overlaid onto the same plot for both $635$ nm (left) and $780$ nm (right) wavelengths. The lower plots show intensity profiles as the beams shift across a row of the array. This plot demonstrates a complete beam waist shift to address five adjacent locations. We measured $>40\%$ system throughput for both wavelengths in the $n=2$ angle multiplication configuration. 

In addition to the system with $v=0.25$ mm, we characterized a system with $v=2.0$ mm to compare the intensity variations between the two systems and the simulation results. For the $v=2.0$ mm, the peak intensity variation among the lattice sites is $<12\%$ for both wavelengths. This matches well with the simulation results. When the two beam system was implemented with $v=0.25$ mm, the peak intensity variations decreased as expected. The simulated intensity variation among the mirror tilt configurations was $3.5\%$, while we saw $<9\%$ for both wavelengths in the system. The descrepancies in simulation and experimental results arise from print-throughs and etch hole features as well as the slight deformation of the MEMS mirror plates during high voltage actuation. While the mirror remains flat at its unactuated state, the strong torsional force and larger spring constant causes the mirror to bow at large tilt angles, leading to peak intensity variations. Improved MEMS mirrors can be designed to address these issues. For an atomic QIP implementation, the variations can be compensated for by altering the duration of the illumination on the trapped atom. 

\begin{figure}[htb]
\centering
		\label{fig:reflectance}
		\includegraphics[width=12cm]{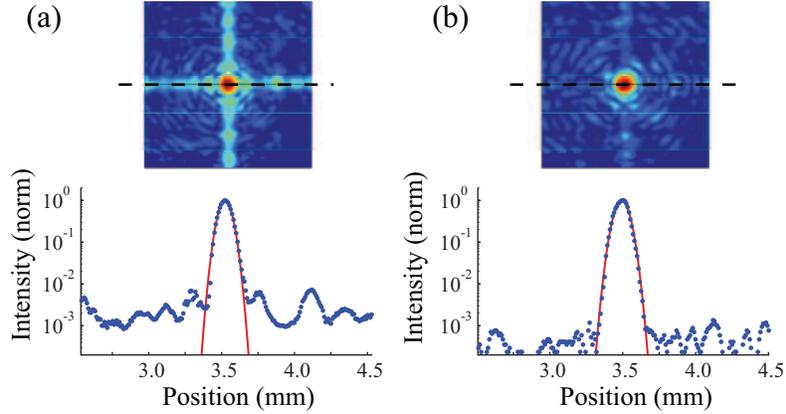}
	\caption{Intensity plots of Gaussian beam at the Fourier plane of a double bounce system. The top plots show a log plot of the intensity data while the bottom plot shows a cross section taken along the black dotted line plotted against the ideal Gaussian shape. (a) is the data taken from a system with a ``D''-shaped electrode and (b) shows data from a ``U''-shaped electrode.}
\end{figure}

To understand how the print-through and etch hole features affect the Gaussian beam, we characterized the quality of the beam at the output of the system and look at the residual intensity at neighboring sites. To this end, we measured the beam profile of a $780$ nm laser beam at the Fourier plane after full propagation through the beam steering system. The waist of the laser beam was $40$ $\mu$m at the MEMS mirrors. Two MEMS mirrors were studied with radius of $100$ $\mu$m and different electrode geometries, ``D''-shaped electrodes and ``U''-shaped electrodes. Each mirror had $4$ etch holes evenly spaced $32$ $\mu$m away from the center of the mirror. The beam profiles were compared with the ideal Gaussian beam shape. Figure \ref{fig:reflectance}(a) shows the intensity data (top) and cross sectional profile (bottom) of the beam for the ``D''-shaped electrode. There is a noticeable diffraction pattern $22$ dB below the peak intensity that is generated by the print-through line traveling down the center of the mirror. On the ``U''-shaped electrode (Fig. \ref{fig:reflectance}(b)), the print-through patterns are moved further from the center of the mirror and the residual intensity at the neighboring lattice sites is $30$ dB below the peak. The effect of the print-through is also seen on the reflectance of the respective mirrors. The ``D''-shaped electrode has a reflectance of $85\%$ while the ``U''-shaped electrode features $90\%$ reflectance. 

\begin{figure}[htb]
\centering
\includegraphics[width=10cm]{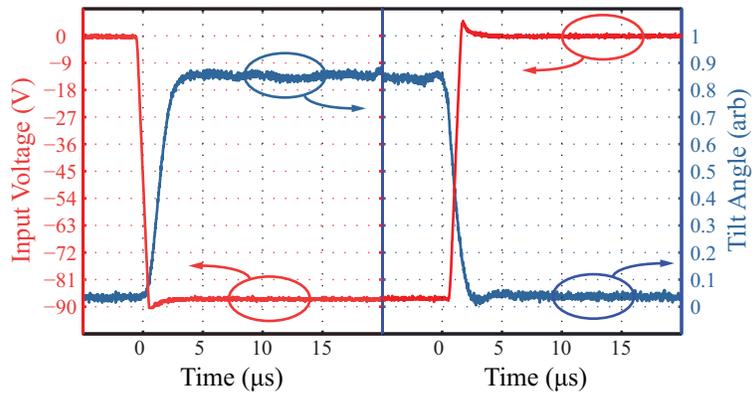}
\caption{\label{fig:trans} Transient response for a $100$ $\mu$m radius system mirror with a resonant frequency of $247$ kHz. Left plot shows ``tilt'' case while the right side shows the ``release'' case. The red line represents the input signal and the blue line indicates the generated tilt angle data from the PSD.}
\end{figure}

We characterized the settling time in our system by measuring the transient characteristics of the MEMS mirrors. Actuating the mirror with a square-wave pulse, we record the displacement of the laser beam deflecting off the mirror on a position sensitive detector (PSD). The mirror shows different damping behaviors under two distinct operating modes. The first (``release'' case) is the case where the mirror relaxes from a tilted to a less tilted position as the magnitude of the applied voltage drops. The second (``tilt'' case) is the case when the applied voltage steps up in magnitude causing the mirror to increase the tilt angle. The mirror's transient response differs in these two cases due to electrostatic softening \cite{ChangsoonSTiQEIJo2007}, resulting in a faster response for the ``release'' case. By designing the mirrors to be slightly underdamped for the ``release'' case, one can achieve optimal settling times for both cases. The result shown in Fig.\ref{fig:trans} demonstrates this with a $100$ $\mu$m radius system mirror with a resonant frequency of $247$ kHz and an angular range of $0.50^{\circ}$. Both the ``tilt'' (left) and ``release'' (right) cases settle in less than $4$ $\mu$s. Further reduction of settling time requires mirrors with larger resonant frequencies and thus a larger spring constant. Due to the limitation on the thickness of the structural layers available in the PolyMUMPs process, it is difficult to reduce the settling time much further without causing the mirror plate to warp during actuation compromising the optical quality of the system. Our simulations indicate that further decrease of settling times down to $1$ $\mu$s will require a change in the fabrication process where thicker springs can be implemented.

\section{Summary}
\label{conclusion}

We have developed a compact, fast optical laser beam steering system capable of handling a broad range of wavelengths and multiple independent beam paths simultaneously. Because the design is a reflection-based scheme that utilizes tilting MEMS mirrors, the system can also provide wavelength multiplexing on a single beam path. System simulations indicate that a $5\times5$ array of positions can be easily addressed with at least $9$ beams simultaneously with better than $12\%$ peak intensity variation across the 2D array for each beam path. Introducing custom optical elements to control Seidel aberrations can reduce the peak intensity variations and increase the number of simultaneous beam paths. We have demonstrated a system that can address a 5$\times$5 array with two independent laser beams ($780$ and $635$ nm) and peak intensity variations across the output array of $9\%$. The system features better than $40\%$ optical throughput, while settling times of $4$ $\mu$s have been measured. Such a system can provide useful functionalities, such as random access control in atomic based quantum information processing.


\section*{Acknowledgments}
\label{acknowledgments}

This work is supported by the Army Research Office under contract W911NF-08-C-0032, and the National Science Foundation under award CCF-0546068.

\end{document}